\documentclass[12pt]{article}
  
\def\fnote#1#2{\begingroup\def\thefootnote{#1}\footnote{#2}\addtocounter
{footnote}{-1}\endgroup}
\usepackage{amsbsy,amssymb,latexsym}
\catcode`@=11
\def\secteqno{\@addtoreset{equation}{section}%
\def\theequation{\thesection.\arabic{equation}}}
\catcode`@=12
\secteqno\newcommand{\be}{\begin{equation}}
\newcommand{\ee}{\end{equation}}
\newcommand{\bea}{\begin{eqnarray}}
\newcommand{\eea}{\end{eqnarray}}
\newcommand{\bref}[1]{(\ref{#1})}
\newcommand{\nn}{\nonumber}

\newcommand{\slX}{/ {\hskip-0.27cm{X}}}
\newcommand{\slY}{/ {\hskip-0.27cm{Y}}}

\newcommand{\slp}{/ {\hskip-0.27cm{p}}}

\newcommand{\slbL}{/ {\hskip-0.27cm{\bf L}}}

\begin{document}
\thispagestyle{empty}
\vfill
\hfill July 17, 2001\par
\hfill  Revised in January 16, 2003\par
\hfill KEK-TH-769\null\par
\hfill DAMTP-2001-50\par
\hfill hep-th/0106114\par
\vskip 20mm
\begin{center}
{\Large\bf Wess-Zumino term for the AdS superstring}\par
\vskip 6mm
{\Large\bf and generalized In\"on\"u-Wigner contraction  }\par
\vskip 6mm
\medskip

\vskip 10mm
{\large Machiko\ Hatsuda~~and~~Makoto\ Sakaguchi
\fnote{$\star$}{
Since October, 2001, Theory division, KEK, e-mail: Makoto.Sakaguchi@kek.jp
}$^\dagger$}\par
\medskip
{\it 
Theory Division,\ High Energy Accelerator Research Organization (KEK),\\
\ Tsukuba,\ Ibaraki,\ 305-0801, Japan} \\
{\it 
$\dagger$ 
Department of Applied Mathematics and Theoretical  Physics\\
Wilberforce Road, Cambridge, CB3 0WA, U.K.}\\
\medskip
E-mail:\ mhatsuda@post.kek.jp,\ 
M.Sakaguchi@damtp.cam.ac.uk
\medskip
\end{center}
\vskip 10mm
\begin{abstract}

\end{abstract} 
We examine a Wess-Zumino term, written in bilinear of superinvariant currents, 
for a superstring in 
 anti-de Sitter (AdS) space. 
The standard In\"on\"u-Wigner contraction 
does not give 
the
 correct flat limit but gives zero. 
This originates from the fact that the fermionic metric of the super-Poincar\'e group is degenerate.
We propose a generalization of the In\"on\"u-Wigner contraction 
which reduces the super-AdS group to the ``nondegenerate" super-Poincar\'e group,
therefore it gives a correct flat limit of this Wess-Zumino term. 
We also discuss the M-algebra obtained 
by this generalized In\"on\"u
-Wigner contraction
from osp(1$\mid$32).\par
\par\par
\newpage
\setcounter{page}{1}
\parskip=7pt
\section{ Introduction}\par
\indent
There has been great interest in studying anti-de Sitter
(AdS)
superstring actions 
\cite{MeTsy,Ram,Berk,HKS,RWS} 
motivated by the AdS/CFT correspondence \cite{Mal}.   
The superstring action contains 
the Wess-Zumino term \cite{GSSUST} which is required by the $\kappa$-symmetry  to match 
the number of dynamical degrees of freedom for bosons and fermions \cite{WSkappa}.
The conventional description of the Wess-Zumino term is used in \cite{MeTsy,Ram},
and an alternative description of the Wess-Zumino term, written in bilinear of superinvariant currents, 
  has been proposed 
for the AdS superstring theories \cite{Berk,RWS}
and for an AdS superstring toy model \cite{HKS}.
The conventional description gives its flat limit straightforwardly as shown in \cite{MeTsy},
while the flat limit of the alternative description
 does not correspond to the simple group contraction.
In 
 flat space the Wess-Zumino term is given as an integral of the closed super-invariant three form
in one dimension higher space \cite{HxM}, 
\bea
S_{WZ}=\int d^{3}\sigma ~  H_3~~,~~H_3=dB_2~~.\label{WZ11}
\eea
The local two form $B_2$ is not super-invariant but pseudo super-invariant.
Its geometrical interpretation given by \cite{azcTow} is that
$H_3$ is an element of the non-trivial class of 
the Chevalley-Eilenberg cohomology for the supertranslation group.
It is also explained that $B_2$ can not be written as the bilinear of the left-invariant (LI) one-forms
 because of the degeneracy of the fermionic metric of
 the supertranslation group \cite{Green}.

In contrast
 to the supertranslation group
the super-AdS group contains a nondegenerate group metric,
so $B_2$ can be written as the bilinear of the LI Cartan one-forms, $L$'s \cite{Berk,RWS}
\bea
S_{WZ}'=\int d^{2}\sigma~B_2~~,~~B_2= g_{\alpha\beta}L^\alpha L^\beta .\label{blWZ}
\eea 
In \cite{HS:PRD},
the bilinear form Wess-Zumino term
for the AdS superstring
has been examined
and it was shown that
the flat limit of the Wess-Zumino term \bref{blWZ}
for the AdS superstring is not obtained 
by the usual In\"on\"u-Wigner (IW) contraction \cite{IW}. 
The IW contraction  does not 
lead to the bilinear form Wess-Zumino term for a flat superstring, but leads to
a total derivative term. 
We consider a  group ``contraction'' such that it leads to the correct flat limit.

The IW contraction relates the super-AdS group and the super-Poincar\'e group as follows.
The super-AdS algebra generated by
${\cal J}_{{\cal A}{\cal B}}=(J_{AB},RP_{A})$ and $\sqrt{R}Q$ is given symbolically as
\bea
\left[ {\cal J}_{\cal AB},{\cal J}_{\cal CD}\right]&=&4\eta_{[{\cal D}|[{\cal A}}{\cal J}_{{\cal B}]|{\cal C}]}\nn\\
\left[ {Q}_{\alpha},{\cal J}_{\cal CD}\right]&=&-\frac{1}{2}({Q}\Gamma_{\cal CD})_\alpha\nn\\
\left\{ {Q}_{\alpha},{Q}_{\beta}\right\}&=&i\eta^{\cal AC}\eta^{\cal BD}
(C\Gamma_{\cal AB})_{\alpha\beta}\frac{1}{R}{\cal J}_{\cal CD}~~,\label{superads}
\eea
where $R$ is the scale parameter representing the radius of the AdS pseudosphere.
For the super-AdS group the Killing supermetric of fermions, 
\footnote{ The Killing supermetric is given by
\bea
\left[
T_M,T_N
\right]=T_Kf^K_{~~MN}~~,~~g_{MN}=f^K_{~~ML}f^L_{~~NK}(-)^{K}\nn~~,
\eea
where $(-)^K=+1(-1)$ if $T_K$ is an even (odd) generator. 
}
\bea
g_{Q_\alpha Q_\beta}=f^{\cal J}_{~~Q_\alpha~Q}f^Q_{~~Q_\beta~ {\cal J}}
			-f^Q_{~~Q_\alpha~ {\cal J}}f^{\cal J}_{~~Q_\beta~Q}
			=i\frac{D(D+1)}{R}C_{\alpha\beta}~~,\label{adsmtr}
\eea
can be chosen 
as the nondegenerate fermionic metric 
to construct the bilinear form Wess-Zumino term \bref{blWZ},
where $C_{\alpha\beta}$ is an antisymmetric charge conjugation matrix.
By the IW contraction 
we obtain the super-Poincar\'e algebra in $1/R\to 0$ limit  
\bea
\begin{array}{cclcccl}
\left[P_A,P_B\right]&=&0&,&\left[J_{AB},J_{CD}\right]&=&4\eta_{[D|[A}J_{B]|C}\\
\left[{\cal Q},P_A\right]&=&0&,&\left[P_A,J_{BC}\right]&=&2\eta_{A[B}P_{C]}\\
\left\{ {\cal Q}_{\alpha},{\cal Q}_{\beta}\right\}&=&2i(C\gamma^A)_{\alpha\beta}P_A&,&
\left[{\cal Q},J_{AB}\right]&=&-\frac{1}{2}{\cal Q}\gamma_{AB}~~.
\end{array}
\label{superpoi}
\eea
The Killing supermetric for the super-Poincar\'e group vanishes
\bea
g_{Q_\alpha Q_\beta}=0~~,
\eea
so $B_2$ can not be written in the bilinear of the LI one-forms in flat space.

It  has been shown that
even in a 
 flat space the bilinear form Wess-Zumino term can be
 constructed from
 the nondegenerate supertranslation group
which contains a
 fermionic center \cite{Green,Siegel}:
\bea
\begin{array}{ccl}
\left[P_A,P_B\right]&=&0\\
\left[Q_\alpha,P_A\right]&=&-\frac{\lambda}{2}(Z\gamma_{A})_\alpha~\\
\left\{ Q_{\alpha}, Q_{\beta}\right\}&=&2i(C\gamma^A)_{\alpha\beta}P_A~~\\
\left[Z_\alpha,P_A\right]&=&\left[Z_\alpha,Q_\beta\right]~=~0~~,
\end{array}
\label{ndsuper}
\eea
where $\lambda$ is a scale parameter.
The nondegenerate group metric can be imposed
\bea
g_{Q_\alpha Z_\beta}=-i\frac{2}{\lambda}C_{\alpha\beta}~~\label{ndm}
\eea
which is not the Killing supermetric, but the invariant group metric 
with
 totally antisymmetric structure constants
\footnote{ 
\bea
f_{Q_\alpha Q_\beta P_A}&=&{\rm Tr}(\{Q_\alpha,Q_\beta\}P_A)
={\rm Tr}([P_A,Q_\alpha]Q_\beta)=f_{P_A Q_\alpha Q_\beta}
=i(C\gamma_A)_{\alpha\beta}\nn\\
{\rm Tr} (P_A P_B)&=&g_{P_A P_B}=\frac{1}{2}\eta_{AB}\nn\\
{\rm Tr} (Q_\alpha Z_\beta)&=&g_{Q_\alpha Z_\beta}=-i\frac{2}{\lambda}C_{\alpha\beta}~~.
\label{ndmetric}
\eea}.
The nondegenerate supertranslation group has been applied to various $p$-brane systems
and the $(p+1)$-form Wess-Zumino terms are used to analyze these systems
\cite{HKS,Sez,SezM,Sakag,AHKT,azc1,Sakag2,HS,nonBPS}.
Now the question becomes what type of group ``contraction'' gives the
nondegenerate super-Poincar\'e group
from the super-AdS group?

The existence of a scale parameter,
$R$ in \bref{superads} and \bref{adsmtr}
and $\lambda$ in \bref{ndsuper} and \bref{ndm},
allows one to have a nondegenerate
metric.
For the usual IW contraction one takes a limit
where such parameter dependence is completely washed out,
as $1/R \to 0$.
As a result the fermionic metric becomes degenerate.
In order to have the bilinear Wess-Zumino term in a flat limit,
equivalently in order to get the nondegenerate super-translation group
in a flat limit,
we need  a
 more general IW contraction 
which keeps some parameter dependence.
We present a generalization of the IW contraction that relates
the super-AdS group and the nondegenerate super-Poincar\'e group
by truncating at 
a
 finite order of the parameter expansion of the LI one-forms.

The organization of this paper is as follows.
In section 2, we propose a generalization of the IW contraction. 
At first we take SO(3) group as a simple example,
and then give an interpretation 
of the generalized IW contraction.
Next we apply this procedure to the super-AdS group,
and 
we show that this generalized IW contraction gives the nondegenerate super-Poincar\'e
group as well as the usual super-Poincar\'e group. 
 In section 3, we discuss the M-algebra \cite{SezM} corresponding to an algebra
 obtained by this generalized IW contraction from osp(1$\mid$32).
The last section devoted to a summary and discussions.

\medskip

The original version of this paper, posted on July 17, 2001,
includes an analysis of the AdS superstring,
which has been published separately in \cite{HS:PRD}.
In this revised version, we focus on
a generalization of the IW contraction,
 which has been discussed further in \cite{dA}.


\section{ Generalized In\"on\"u-Wigner contraction}\par
\indent

It was shown \cite{HS:PRD} that the flat limit of 
the bilinear form Wess-Zumino term 
must be taken in such a way that one 
keeps
not only leading terms but also next to leading terms 
of Cartan one-forms in the $1/R$ expansion.
The usual IW contraction reduces the super-AdS algebra to
the super-Poincar\'e algebra and keeps only leading terms,
therefore the flat limit action becomes the total derivative part. 
In order to keep the next to leading term,
we need to enlarge the ``contraction'' procedure.
In this section we will consider a generalization of the IW contraction 
in such a way that it keeps the next to leading term and 
it also includes the original IW contraction. 
At first we take a simple example, SO(3) which is reduced to ISO(2) with a center
by the generalization of the IW contraction. Next we will show that the super-AdS group is reduced to the 
``nondegenerate" super-Poincar\'e group by the generalization.    
\par

\subsection{ Generalized contraction of SO(3)}
\par
We begin with SO(3) as the simple example of a generalization of the IW contraction.
Generators $J_I$, $I=(x,y,z)$ satisfy 
\bea
\left[J_I,J_J \right]=\epsilon_{IJK}J_K\label{xyz}~~.
\eea
Parameterizing a group element as $g=e^{J_I\phi^I}$ 
gives Cartan
one-forms
\bea
L^I&=&d\phi^I+\frac{1}{2}d\phi^J\phi^K\epsilon^{IJK}
+\frac{1}{3!}d\phi^J(\phi^I\phi^J-\delta^{IJ}\phi^2)
-\frac{1}{4!}d\phi^J\phi^K\epsilon^{IJK}\phi^2+\cdots\label{Lphi}~~,
\eea
 defined by $g^{-1}dg=J_IL^I$ and satisfying
the Maurer-Cartan (MC)
 equation, $dL^I=\frac{1}{2}\epsilon^{IJK}L^JL^K$.
The conventional IW contraction requires rescaling of generators
\bea
J_x\to \frac{1}{s}J_x~~,~~J_y\to \frac{1}{s}J_y~~,~~J_z\to J_z~~,\label{sss}
\eea
and taking the limit $s\to 0$.
This corresponds to rescaling parameters,
\bea
&&\phi^x\to {s}\phi^x~~,~~\phi^y\to {s}\phi^y~~,~~\phi^z\to \phi^z~~,\label{ssphi}
\eea
and $L^I$'s,  $L^x\to {s}L^x~,~L^y\to {s}L^y~,~L^z\to L^z$, and then taking the $s\to 0$ limit.
This procedure is equivalent to keeping only leading terms of $L^I$'s in $s$-expansion.

Now, before taking the limit, we consider the scaling \bref{ssphi} for $\phi^I$'s which implies 
the following expansion of Cartan 
one-forms $L^I$'s 
\bea
L^x&=&sL^x_1+s^3L^x_3+\cdots=\bigoplus_{i=0}^\infty s^{2i+1}L^x_{2i+1}\nn\\
L^y&=&sL^y_1+s^3L^y_3+\cdots=\bigoplus_{i=0}^\infty s^{2i+1}L^y_{2i+1}\label{LsLs}\\
L^z&=&L^z_0+s^2L^z_2+\cdots=\bigoplus_{j=0}^\infty s^{2j}L^z_{2j}\nn
\eea
where
\bea
L^x_1&=&d\phi^x+\frac{1}{2}(d\phi^y\phi^z-d\phi^z\phi^y)+\cdots,~~\cdots\nn\\
L^y_1&=&d\phi^y+\frac{1}{2}(d\phi^z\phi^x-d\phi^x\phi^z)+\cdots,~~\cdots\nn\\
L^z_0&=&d\phi^z~~,~~\label{L012}\\
L^z_2&=&\frac{1}{2}(d\phi^x\phi^y-d\phi^y\phi^x)+
\frac{1}{3!}\{(d\phi^x\phi^z-d\phi^z\phi^x)\phi^x
		+(d\phi^y\phi^z-d\phi^z\phi^y)\phi^y\}+\cdots,~~\nn\\
\cdots&&~~\nn
\eea
MC equations are satisfied in order by order
\bea
dL^I_M=\frac{1}{2}\epsilon^{IJK}L^J_NL^K_{M-N}\label{obo}
\eea
or equivalently 
\bea
dL^z_0&=&0\label{SO1}\\
dL^x_1&=&L_1^yL_0^z\label{SO2}\\
dL^y_1&=&L_0^zL_1^x\label{SO3}\\
dL^z_2&=&L_1^xL_1^y\label{SO4}\\
\cdots&&\nn~~.
\eea
The conventional IW contraction keeps only MC equations \bref{SO1}-\bref{SO3}
for $L^x_1,L^y_1,L^z_0$, and corresponding algebra is
\bea
\left[J_{x,1},J_{y,1}\right]&=&0\nn\\
\left[J_{y,1},J_{z,0}\right]&=&J_{x,1}\nn\\
\left[J_{z,0},J_{x,1}\right]&=&J_{y,1}\label{ISO2}~~.
\eea
This is the inhomogeneous SO(2); 
the rotation around the $z$-axis on the $x$-$y$ plain remains, but rotations around the $x$-axis 
and $y$-axis are restricted to infinitesimal values and turn out to be 
translations in the $y$ and $x$ directions.

Next let us consider alternative limit in such a way that
not only leading terms, $L^x_1,L^y_1,L^z_0$, but also the next to leading term
, $L^z_2$, are kept.
Then survived MC equations are \bref{SO1}-\bref{SO4},
and corresponding algebra is
\bea
\left[J_{x,1},J_{y,1}\right]&=&J_{z,2}\nn\\
\left[J_{y,1},J_{z,0}\right]&=&J_{x,1}\nn\\
\left[J_{z,0},J_{x,1}\right]&=&J_{y,1}\label{ISO22}~~.
\eea
The new generator, $J_{z,2}$, is a 
central term.
This is an intermediate form between
the three-dimensional rotation on a two-dimensional sphere 
and
the rotation of the $z$-axis on a $x$-$y$ flat plane.
Therefore this generalization brings the rotation of the $z$-axis on the $x$-$y$ plane
which has a infinitesimally small curvature. 
Further generalization is possible as long as the
 MC equations are
consistently maintained.
\par

\subsection{ Generalized contraction of super-AdS}
\par
Now let us go back to the problem of the Wess-Zumino term for a AdS superstring.
The usual IW contraction reduces 
the super-AdS group \bref{superads} to the super-Poincar\'e group
\bref{superpoi}.
The MC equations for the super-AdS group\footnote{
We follow the notation used in \cite{MeTsy};
$A=(a,a')$~($a=0,1,...,4,~a'=5,6,...,9$),
$\alpha=\alpha'\alpha''$ ($\alpha'=1,...,4,~\alpha''=1,...,4)$, and
$I=1,2$.
} are 
\bea
DL^{\alpha I}&=&\epsilon^{IJ}\frac{i}{2}(\slbL L^{J})^\alpha\nn\\
D{\bf L}^A&=&i\bar{L}\gamma^A L\label{SAdSMC}\\
D{\bf L}^{AB}&=&-\frac{1}{2}{\bf L}^A{\bf L}^B-\frac{1}{2}\bar{L}^I\gamma^{AB}\epsilon^{IJ} L^J
~~\nn
\eea
where covariant derivative $D$ includes the Lorentz rotation.
They are  reduced under the IW contraction to those for
the super-Poincar\'e algebra
\footnote{  
Although the Cartan one-form for the Lorentz generator 
vanishes for the super-Poincar\'e algebra,
we include it in order to clarify the commutation relations of the corresponding superalgebra.
}
\bea
DL^{\alpha I}&=&0\nn\\
D{\bf L}^A&=&
i\bar{L}\gamma^A L\label{after}\\
D{\bf L}^{AB}&=&0\label{SPMC}~~.\nn
\eea

\medskip
The procedure for keeping
 the next to leading term of Cartan one-forms 
is as follows.
We expand the Cartan one-forms in $s=1/R$ as
\bea
{\bf L}^{A}=\displaystyle\bigoplus_{i=1}^\infty s^{i} {\bf L}^{A}_{i}~~,\label{expRLL}~~
{\bf L}^{AB}=\displaystyle\bigoplus_{j=0}^\infty s^{j} {\bf \Omega}^{AB}_{j}~~,\label{expAB}
~~
L^{I\alpha}=\displaystyle\bigoplus_{k=0}^\infty s^{k+1/2} L^{I\alpha}_{k+1/2}
\eea
and then the MC equations are satisfied in each order of $R$
\bea
D{\bf \Omega}^{AB}_{0}&=&0\label{1}\\
DL^{I\alpha}_{1/2}&=&0\label{2}\\
D{\bf L}^A_{1}&=&i\bar{L}^{I}_{1/2}\gamma^A L^I_{1/2}\label{3}\\
D{\bf \Omega}^{AB}_{1}&=&-i\bar{L}^{I}_{1/2}\epsilon_{IJ}\gamma^{AB} L^J_{1/2}\label{4}\\
DL^{I\alpha}_{3/2}&=&\epsilon^{IJ}\frac{i}{2}\slbL_{1} L^{J}_{1/2}\label{5}\\
&\vdots &\nn
\eea
where $D$ includes Lorentz rotation.
The equations \bref{1}-\bref{3} correspond to 
the super-Poincar\'e algebra \bref{after},
in other words, 
leading terms of this expansion correspond to
the usual IW contraction.

Now we generalize the IW contraction as follows.
For an expansion parameter $s$ which is finite before contraction,
we expand an element and subtract 
the redundant part,
\bea
T(s)&=&\displaystyle\bigoplus_{n=0}^\infty s^n T_n\nn\\
&&\nn\\
~\to~T(s)&=&T(s)-\displaystyle\int_0^s~d^{N+1}s' \left(\frac{d}{d{s'}}\right)^{N+1}T(s')\label{gencntr}\\
&=&T_0+sT_1+\cdot\cdot\cdot+s^NT_N~~.\nn
\eea
The Cartan one-forms \bref{expRLL} 
are now subtracted, 
\begin{eqnarray}
{\bf L}^{AB}
&\to&
\bigoplus_{j=0}^{N} s^{j} {\bf \Omega}_{j}^{AB}
=  {\bf L}^{AB}-\int_0^sd^{N+1}s\partial_s^{N+1} {\bf L}^{AB}~~,\nn\\
{\bf L}^A
&\to&\bigoplus_{i=1}^{M} s^i {\bf L}_{i}^A
=  {\bf L}^A-\int_0^sd^{M+1}s\partial_s^{M+1} {\bf L}^A
,\label{gradeLL}\label{grade}\\
  L^\alpha&\to&
 \bigoplus_{k=0}^{P} s^{k+1/2} L_{k+1/2}^{\alpha}
= L^\alpha-s^{-1/2}\int_0^sd^{P+1}s \partial_s^{P+1} (s^{1/2} L^\alpha)\nn~~.
\end{eqnarray}
Taking
$(N,M,P+\frac{1}{2})$ to be $(1,1,\frac{3}{2})$
gives LI Cartan one-forms 
$({\bf \Omega}_0,{\bf \Omega}_1,{\bf L}_1,L_{1/2},  L_{3/2})$ 
satisfying \bref{1}-\bref{5}. 
This procedure can give the correct flat limit of the AdS superstring.
In this subsection we focus on a case $N=1,M=1,P=1$, while  more
 general cases will be discussed in the next subsection.

Let us translate  the
 MC equations \bref{1}-\bref{5} into (anti)commutation relations of nondegenerate algebra.
If we expand the LI Cartan one-form introducing generators as
\footnote{We omit the Lorentz generator here.
This implies that the commutation relations
corresponding to Lorentz transformations are omitted for simplicity below.}
\bea
G^{-1}dG=L_{1/2} Q_{1/2}+L_{3/2} Q_{3/2}+{\bf L}_1P_1+\frac{1}{2}{\bf \Omega}_1J_{1},
\eea 
MC equations \bref{1}-\bref{5} turn out to be the nondegenerate superalgebra
on $\Bbb E^{1,4}\times  \Bbb E^{5}$
\bea
\left\{Q_{I,1/2},Q_{J,1/2}\right\}&=&-2i\delta_{IJ}(C\gamma^A)P_{A,1}
	+\epsilon_{IJ}(C\gamma^{AB})J_{AB,1}\nn\\
\left[Q_{I,1/2},P_{A,1}\right]&=&\frac{i}{2}Q_{J,3/2}\gamma_A\epsilon_{JI}\label{QPQ}\\
\left[Q_{I,1/2},J_{AB,1}\right]&=&-\frac{1}{2}Q_{I,3/2}\gamma_{AB}\nn~~\\
{\rm others}&=&0~~,\nn
\eea
where the Jacobi identity of three $Q$'s
requires 
a tensor central charge which can be always taken as a
quotient subgroup.
We have obtained the nondegenerate superalgebra from super-AdS algebra 
using generalized IW contraction.
This algebra can be rewritten in 10-dimensional spinor representation\footnote{  
In the limit $R\to \infty$, we have obtained superalgebras on
$\Bbb E^{1,4}\times \Bbb E^{5}$.
In order to obtain $\Bbb E^{1,9}$ superalgebra,
we must recover full generators of the ten-dimensional covariance.
This is associated with the replacement
of the $\Bbb E^{1,4}\times \Bbb E^{5}$ Fierz identity in the Metsaev-Tseytlin notation
\begin{eqnarray}
 (\gamma_A)^{I\alpha}{}_{(J\beta}
  (C\gamma^A)_{K\gamma ~L\delta)}
 +\frac{1}{2}(\epsilon\gamma_{AB})^{I\alpha}{}_{(J\beta}
  (\epsilon C\gamma^{AB})_{K\gamma ~L\delta)}=0
\end{eqnarray}
with
the $\Bbb E^{1,9}$ Fierz identity
\begin{eqnarray}
(\delta\gamma_{A})^{I\alpha}{}_{(J\beta}(\delta\gamma^A)_{K\gamma~L\delta )}
+(\tau_3\gamma_{A})^{I\alpha}{}_{(J\beta}(\tau_3\gamma^A)_{K\gamma~L\delta )}
=0.
\end{eqnarray}
In this particular case, $J_1$ is replaced with the F-string charge
 ($\Sigma_A$) in the algebra.
}
 as
\bea
\left\{Q_{I,1/2},Q_{J,1/2}\right\}&=&-2i\delta_{IJ}(C\gamma^A)P_{A,1}
	+(\tau_3)_{IJ}(C\gamma^{A})\Sigma_{A,1}\nn\\
\left[Q_{1,1/2},(P_{A,1}+\Sigma_{A,1})\right]&=&\frac{i}{2}Q_{1,3/2}\gamma_A\label{QPQ2}\\
\left[Q_{2,1/2},(P_{A,1}-\Sigma_{A,1})\right]&=&\frac{i}{2}Q_{2,3/2}\gamma_A\label{QPQ3}\nn\\
{\rm others }&=&0~~,\nn
\eea 
where the following cyclic identity
\bea
(C\gamma_A\delta)^{I\alpha}_{(J\beta}(C\gamma^A\delta)_{K\gamma~L\delta)}
+(C\gamma_A\tau_3)^{I\alpha}_{(J\beta}(C\gamma^A\tau_3)_{K\gamma~L\delta)}
=0~~
\eea
is used
and the center $\Sigma_A$ can be taken as a quotient subgroup.
Of course it is necessary to check the cyclic identity.
This procedure may correspond to the similarity transformation 
in the original IW contraction 
which is required for the consistency.

There are two manners 
for
 realizing these algebras:
\par\noindent
(i) A simple example of realization of the algebra \bref{QPQ2} is that for
an open superstring 
as discussed in \cite{HS},
\bea
Q_{1/2}&=&\int~[\zeta-i\bar{\theta}(\slp+T\tau_3\slX')-\frac{iT}{6}
(\bar{\theta}\gamma \tau_3\theta'\cdot \bar{\theta}\gamma
+\bar{\theta}\gamma \theta'\cdot \bar{\theta}\gamma\tau_3)]
\nn\\
Q_{3/2}&=&\int~T\bar{\theta}'\label{opsust}\\
P_{1}&=&\int~p~~,~~\Sigma~=~T\int~X'\nn~~.
\eea
The system is described by only $(X^A,p_A)$ and $(\theta^\alpha,\zeta_\alpha)$
\footnote{  
The two-form doublet $B$ are written in terms of Cartan one-forms as
\cite{Sakag,AHKT,Sakag2}
\begin{eqnarray}
B_{NS}&\equiv& \bar L_{1/2}\sigma_3 L_{3/2}-\frac{1}{2}{\bf \Omega}_{NS,1}{\bf L}_1,\qquad
dB_{NS}=i{\bf L}_1(\bar L_{1/2}\sigma_3\gamma L_{1/2})=H_{NS},
\nn\\
B_{RR}&\equiv& \bar L_{1/2}\sigma_1 L_{3/2}-\frac{1}{2}{\bf \Omega}_{RR,1}{\bf L}_1,\qquad
dB_{RR}=i{\bf L}_1(\bar L_{1/2}\sigma_1\gamma L_{1/2})=H_{RR}.
\nn\\
\end{eqnarray}
Parameterizing a group element as $G=e^{x^AP_A}e^{\theta Q_{1/2}}$,
Cartan one-forms are expressed as
\begin{eqnarray}
G^{-1}dG&=&{\bf L}^AP_A+{\bf \Omega}^A\Sigma_A+L_{1/2}Q_{1/2}+L_{3/2}Q_{3/2},\nonumber\\
{\bf L}^A&=&dx^A-i(\bar\theta\gamma^Ad\theta),\nonumber\\
{\bf \Omega}^A_{NS}&=&-i(\bar\theta\sigma_3\gamma^Ad\theta),\nonumber\\
{\bf \Omega}^A_{RR}&=&-i(\bar\theta\sigma_1\gamma^Ad\theta),\nonumber\\
L_{1/2}&=&d\theta,\nonumber\\
L_{3/2}&=&-\frac{i}{2}dx^A(\gamma_A\theta)^\alpha
  -\frac{1}{6}(\bar\theta\gamma d\theta)(\gamma\theta)^\alpha
  -\frac{1}{6}(\bar\theta\sigma_3\gamma d\theta)(\sigma_3\gamma\theta)^\alpha,\nonumber
\end{eqnarray}
and $B=B_{NS}$ is expressed in terms of $(x,\theta)$
\begin{eqnarray}
B=-idx(\bar\theta\sigma_3\gamma d\theta)
  -\frac{1}{2}(\bar\theta\gamma d\theta)(\bar\theta\sigma_3\gamma d\theta).
\nn\label{NSNS2}
\end{eqnarray}
This agrees with the two-form $B$ obtained in (2.26). For RR two-form, $\sigma_3$ in \bref{NSNS2}
is replaced by
$\sigma_1$.}.
This realization can be viewed as a description of the nondegenerate group manifold 
in terms of the AdS superspace.

\par\noindent
(ii) Another way is to introduce canonical coordinates $(\xi^\alpha, \Pi_{\xi,\alpha})$
and $(y^A,p_Y^A)$ corresponding to $Q_{\alpha,3/2}$ and $\Sigma_A$ adding to the original variables.
This is the case in our previous work \cite{HKS,nonBPS}.
The canonical generators are expressed as
\bea
Q_{1/2}&=&\int ~[\zeta-i\bar{\theta}(\slp+\slp_Y\tau_3)
-\Pi_\xi(\slX+\slY\tau_3)
-\frac{i}{3}
(\Pi_\xi \gamma\theta\cdot\bar{\theta}\gamma+
\Pi_\xi \gamma\tau_3\theta\cdot\bar{\theta}\gamma\tau_3)]\nn\\
Q_{3/2}&=&\int~\Pi_\xi\label{Q12}\\
P_{1}&=&\int~p~~,~~\Sigma~=~\int~p_Y\nn
\eea
satisfying \bref{QPQ2}.

The algebra \bref{QPQ2} gives the group metric such that 
the metric in fermionic coordinates becomes nondegenerate,
$g_{Q_{1/2}Q_{3/2}}={\rm tr}(Q_{1/2}Q_{3/2})\neq 0$. 
The nondegenerate group manifold is obtained by the generalized IW contraction
from the AdS superspace in these ways.

\medskip

\subsection{ Sequence of the generalized IW contracted algebras }\par
\indent

Now let us consider the general integer $N,M,P$ case in \bref{grade}.
A set of generators with
the highest dimension $(N, M, P+1/2)$
form an algebra ${\cal G}(N, M, P+1/2)$
\bea
D\Omega_j^{AB}&=&-\sum_{l=1}^{j-1}
  \frac{1}{2}{\bf L}_l^A{\bf L}_{j-l}^B
  -\sum_{l=0}^{j-1}
  \frac{1}{2}\bar{L}_{l+1/2}^I\gamma^{AB}\epsilon^{IJ} L^J_{j-l-1/2}~~,
~~~~~j=0,...,N~~,
~~\nn
\\
D{\bf L}_{i}^A&=&\sum_{l=0}^{i-1}
  i\bar{L}_{l+1/2}\gamma^A L_{i-l-1/2}~~,
~~~~~i=1,...,M~,\label{g(N,M,P+1/2)}\\
DL_{k+1/2}^{\alpha I}&=&\sum_{l=1}^k
  \epsilon^{IJ}\frac{i}{2}\textbf{L}_l^A(\gamma_A L_{k-l+1/2})^\alpha~~,
~~~~~k=0,...,P~,\nn
\eea
with
$N=M=P$, $N=M=P+1$, $N=M+1=P+1$ or $N+1=M=P+1$.
For example, ${\cal G}(1,1,3/2)$ is the nondegenerate superalgebra
generated by
$\{{\bf \Omega}_0, {\bf \Omega}_1, {\bf L}_1, L_{1/2}, L_{3/2}\}$
discussed in the previous subsection.
A generator with the highest dimension 
is a center of the algebra.
In other words, this algebra is nilpotent.
This implies a sequence of algebras
\begin{eqnarray}
{\cal G}(0,0,0)
\to  {\cal G}(0,0,\frac{1}{2})
\to \left\{
  \begin{array}{l}
    {\cal G}(0,1,\frac{1}{2})   \\
    {\cal G}(1,0,\frac{1}{2})   \\
  \end{array}
\right.
\to {\cal G}(1,1,\frac{1}{2})
\to {\cal G}(1,1,\frac{3}{2})
\to \cdots 
\nn
\end{eqnarray}
\bea
&&~~ {\cal G}(0,0,0)=\{{\bf \Omega}_{0}\}:{\rm Lorentz}~ {\rm algebra} \nn\\
&&~~ {\cal G}(0,1,0)=\{{\bf \Omega}_{0},{\bf L}_{1}\}:\mbox{Poincar\'e}~{\rm algebra} \nn\\ 
&&~~{\cal G}(0,1,1/2)=\{{\bf \Omega}_{0},{\bf L}_{1},L_{1/2}\}:{\rm the~ (degenerate)~ super}\mbox{-Poincar\'e}~{\rm algebra}\nn\\
&&~~{\cal G}(1,1,3/2)=\{{\bf \Omega}_{0},{\bf \Omega}_{1},{\bf L}_{1},L_{1/2},L_{3/2}\}:{\rm the~
 nondegenerate~ super}\mbox{-Poincar\'e}~ {\rm algebra}\nn
\eea
where an arrow indicates an central extension.
${\cal G}(\infty)$ is the original super-AdS algebra after suitable combination of generators.

As well as the nondegenerate supertranslation algebra case \bref{QPQ}, 
there are two manners of describing group manifold of ${\cal G}(N,M,P+\frac{1}{2})$.\par\noindent
(i)The first one is to begin with a super-AdS group manifold
and to regard ${\cal G}(N,M,P+\frac{1}{2})$ as a corresponding limit as we have seen above.
In this case, the group manifold of ${\cal G}(N,M,P+\frac{1}{2})$ is described by AdS supercoordinates
$(x,\theta)$ dual to $(P,Q)$ (after dividing by Lorentz group)
with nontrivial boundary conditions.\par\noindent
(ii)The other one is to introduce a coordinate set
dual to generators with dimension from $0$ to $k$.
The group manifold is parameterized by coordinates dual to them.
\par

\medskip

\section{ M-algebra and the contracted osp(1$\mid$32)}\par
\indent

There is another interesting example of this generalized IW contraction. 
M-algebra \cite{SezM} is an eleven dimensional superalgebra
generated by generators of supertranslations $Q_M=(Q_\alpha, P_\mu)$,
string charge $Z_M$, M2-brane charges $Z_{MN}$ and M5-brane charges $Z_{M_1\cdots M_5}$,
where $M$ runs 11-dimensional vector indices and spinor indices 
   $M=(\mu,\alpha)$.
This algebra is too big enough to include the D=11 N=1 supertranslation algebra. 
In order to examine the structure of the M-algebra,
we rather begin with osp(1$\mid$32) algebra
which is generated by
 $32$ supercharges ${\cal Q}_\alpha
 $
and $528$ sp(32) generators ${\cal M}_{\alpha\beta}
$, $\alpha,\beta=1,\ldots ,32$ 
in spinor indices.
They satisfy the following algebra
\begin{eqnarray}
\{{\cal Q}_\alpha, {\cal Q}_\beta\}={\cal M}_{\alpha\beta},\quad
[{\cal Q}_\alpha, {\cal M}_{\beta\gamma}]=\Omega_{\alpha\beta}{\cal Q}_\gamma,\quad
[{\cal M}_{\alpha\beta},{\cal M}_{\gamma\delta}]=\Omega_{\beta\gamma}{\cal M}_{\alpha\delta}.
\end{eqnarray}
We denote Cartan one-forms
 dual to $\cal Q_\alpha$, $\cal M_{\alpha\beta}$
as $\Pi^\alpha$ and $\bf{\Pi}^{\alpha\beta}$, respectively and expand as
\bea
\Pi^{\alpha}&=&\displaystyle\bigoplus_{n=0} \frac{1}{R^{(n+1/2)}}\Pi^{\alpha}_{n+1/2}\nn\\
{\bf \Pi}^{\alpha\beta}&=&\displaystyle\bigoplus_{n=1} \frac{1}{R^{n}}{\bf \Pi}^{\alpha\beta}_{n}\label{Rexp}~~.
\eea
Cartan one-forms
and the dimensions are listed below.
\begin{table}[htbp]
 \begin{center}
  \begin{tabular}{|c|c|c|c|c|c|c|c|}
    \hline
     generator  &  $Q_\alpha$  & $P_{\alpha\beta}$   & $Z_\alpha$   & $M_{\alpha\beta}^{(2)}$   & $Q_\alpha^{(5/2)}$   & $M_{\alpha\beta}^{(3)}$   & $Q_\alpha^{(7/2)}$   \\
    \hline
     dimension  & 1/2   & 1   & 3/2   & 2   &  5/2  & 3   & 7/2   \\
    \hline
    Cartan one-form
       & $\Pi^\alpha$   &  $\Pi^{\alpha\beta}$  & $\Pi^\alpha_{3/2}$   & $\Pi^{\alpha\beta}_{2}$   &  $\Pi^{\alpha}_{5/2}$  & $\Pi^{\alpha\beta}_{3}$   & $\Pi^\alpha_{7/2}$   \\
    \hline
  \end{tabular}
 \end{center}
\end{table}\par\noindent
A general
 expansion gives infinite number of generators,
but we terminate this sequence at the suitable dimensions in order to focus on the M-algebra.

 The
MC equations for osp(1$\mid$32) in each order of $R$ are found to be
\begin{eqnarray}
d\Pi^{\alpha}&=&0,\label{11}\nn\\
d\Pi^{\alpha\beta}&=&
  \Pi^\alpha\Pi^\beta, \nn\\
d\Pi^{\alpha}_{3/2}&=&
  \Omega_{\beta\gamma}\Pi^{\alpha\beta}\Pi^{\gamma},\nn\\
d\Pi^{\alpha\beta}_{2}&=&
  \Pi^\alpha\Pi^\beta_{3/2}
  +\frac{1}{2}\Omega_{\gamma\delta}\Pi^{\gamma\alpha}\Pi^{\beta\delta}, \\
d\Pi^{\alpha}_{5/2}&=&
  \Omega_{\beta\gamma}\Pi^{\alpha\beta}_{2}\Pi^{\gamma}
  +\frac{1}{2}\Omega_{\beta\gamma}\Pi^{\alpha\beta}\Pi^{\gamma}_{3/2}, \nn\\
d\Pi^{\alpha\beta}_{3}&=&
  \Pi^{\alpha}\Pi^{\beta}_{5/2}
  +\Omega_{\gamma\delta}\Pi^{\gamma\alpha}\Pi^{\beta\delta}_{2}
  +\frac{1}{4}\Pi^{\alpha}_{3/2}\Pi^{\beta}_{3/2}, \nn\\
d\Pi^{\alpha}_{7/2}&=&
  \Omega_{\beta\gamma}\Pi^{\alpha\beta}_{3}\Pi^{\gamma}
  +\frac{1}{2}\Omega_{\beta\gamma}\Pi^{\alpha\beta}\Pi^{\gamma}_{5/2}
  +\frac{1}{2}\Omega_{\beta\gamma}\Pi^{\alpha\beta}_{2}\Pi^{\gamma}_{3/2}.\label{17}\nn
\end{eqnarray}
This corresponds to
 the M-algebra proposed by 
Sezgin\cite{SezM},
if we relate our Cartan one-forms with his Cartan one-forms, $e^{M}$, $e'_{M}$, $e_{MN}$,
 $e_{M_1\cdots M_5}$, which are dual to $Q_M$, $Z_M$, $Z_{MN}$ and $Z_{M_1\cdots M_5}$ as
\bea
\Pi^\alpha&=&e^\alpha\nn\\
 \Pi^{\alpha\beta}&=& 
 (\gamma^\mu)_{\alpha\beta} e'_{\mu}+(\gamma_\mu)_{\alpha\beta} e^{\mu}
 +(\gamma^{\mu\nu})_{\alpha\beta} e_{\mu\nu}+(\gamma^{\mu_1 \cdots \mu_5})_{\alpha\beta} e_{\mu_1 \cdots \mu_5}\nn\\
 \Pi^\alpha_{3/2}&=&e'_\alpha+  (\gamma^\mu)_{\beta\alpha} e_{\mu\beta}
  +(\gamma^{\mu_1\cdots \mu_4})_\beta{}_\alpha e_{\mu_1\cdots \mu_4\beta}  \nn\\
 { \Pi}^{\alpha\beta}_{2} &=&e_{\alpha\beta}+(C\gamma^{\mu\nu\rho})^\alpha_{\ \gamma} e_{\mu\nu\rho\gamma\beta}\label{Pie}
\eea
etc.,
or equivalently
\begin{eqnarray}
Q_\alpha&=&Q_\alpha\nn\\
P_{\alpha\beta}&=&
  (C\gamma^\mu)_{\alpha\beta}P_\mu
  + (C\gamma_\mu)_{\alpha\beta}Z^\mu
  +\frac{1}{2!}(C\gamma_{\mu\nu})^{\alpha\beta}Z^{\mu\nu}
  +\frac{1}{5!}(C\gamma_{\mu_1\cdots \mu_5})_{\alpha\beta}Z^{\mu_1\cdots \mu_5},
  \nonumber\\
Z_\alpha&=&Z^\alpha+
  (\gamma_\mu)_\beta{}_\alpha Z^{\mu\beta}
  +(\gamma_{\mu_1\cdots \mu_4})_{\alpha\beta} Z^{\mu_1\cdots \mu_4\beta} \nn\\
M_{\alpha\beta}^{(2)}&=&Z^{\alpha\beta}+(C\gamma_{\mu\nu\rho})_{\alpha\gamma}Z^{\mu\nu\rho\gamma\beta}\label{ospm2}
\end{eqnarray}
etc.
In Sezgin's M-algebra the
 auxiliary generators $Z$ with spinor indices, $Z^{\mu\beta},Z^{\alpha\beta}$ etc.,
correspond to the higher dimension generators, $Z_\alpha,M^{(2)}_{\alpha\beta}$ etc., in our algebra.
Our reduced M-algebra is subalgebra of 
 Sezgin's M-algebra;
the number of bosonic generators and fermionic generators are
$528\times 3=1584$ and $32\times 4=128$ for ours and
$664147$
\footnote{  
$\left(\!\!
  \begin{array}{c}
    11   \\
     1  \\
  \end{array}
 \!\!\right)
+\left(\!\!
  \begin{array}{c}
    11   \\
     1  \\
  \end{array}
 \!\!\right)
+\left(\!\!\!
  \begin{array}{c}
    11   \\
    2   \\
  \end{array}
 \!\!\!\right)
+{\left(\!\!
  \begin{array}{c}
    32   \\
     2  \\
  \end{array}
 \!\!\right)_{sym}}
+\left(\!\!\!
  \begin{array}{c}
    11   \\
     5  \\
  \end{array}
 \!\!\!\right)
 +\left(\!\!
  \begin{array}{c}
    11   \\
     3  \\
  \end{array}
 \!\!\right)\times
 \left(\!\!
  \begin{array}{c}
    32   \\
     2  \\
  \end{array}
 \!\!\right)_{sym}
+\left(\!\!
  \begin{array}{c}
    11   \\
     1  \\
  \end{array}
 \!\!\right)
 \times
\left(\!\!
  \begin{array}{c}
    32   \\
     4  \\
  \end{array}
 \!\!\right)_{sym}
={664147}.
$
}
 and 
$717088$  
\footnote{
$
\left(\!\!
  \begin{array}{c}
    32   \\
     1  \\
  \end{array}
 \!\!\right)
+\left(\!\!
  \begin{array}{c}
    32   \\
     1  \\
  \end{array}
 \!\!\right)
+\left(\!\!
  \begin{array}{c}
    32   \\
     1  \\
  \end{array}
 \!\!\right)
 \times
\left(\!\!
  \begin{array}{c}
    11   \\
     1  \\
  \end{array}
 \!\!\right)
 +\left(\!\!
  \begin{array}{c}
    11   \\
     4  \\
  \end{array}
 \!\!\right)
 \times
 \left(\!\!
  \begin{array}{c}
    32   \\
     1  \\
  \end{array}
 \!\!\right)
+\left(\!\!
  \begin{array}{c}
    11   \\
     2  \\
  \end{array}
 \!\!\right)
\times \left(\!\!
  \begin{array}{c}
    32   \\
     3  \\
  \end{array}
 \!\!\right)_{sym}+\left(\!\!
  \begin{array}{c}
    32   \\
     5  \\
  \end{array}
 \!\!\right)_{sym}=717088
\nn.
$ }
for 
 Sezgin's M-algebra, 
 respectively.

A set of generators with dimension less than $k$ forms an algebra. 
We denote this as ${\cal G}(k)$.
Then this reduced M-algebra is ${\cal G}(7/2)$. 
Generator with the highest dimension is a center of the algebra.
This algebra contains algebras
\begin{eqnarray}
&&{\cal G}(1)\to{\cal G}(3/2)\to
\cdots \to{\cal G}(7/2)\nn\\
&&~~{\cal G}(1):{\rm super}\mbox{-}{\rm translation~algebra~with~maximal~central~extension}\nn\\
&&~~{\cal G}(3/2):{\rm nondegenerate~algebra}\nn\\
&&~~{\cal G}(7/2):{\rm reduced}~{\rm M}\mbox{-}{\rm algebra}\nn~~.
\end{eqnarray}
\par
\medskip

One can extend the reduced M-algebra including a dimension $4$ generator
or a Cartan 
one-form $\Pi^{\alpha\beta}_4$ satisfying the MC equation
\begin{eqnarray}
d\Pi^{\alpha\beta}_{4}=
  \Pi^{\alpha}\Pi^{\beta}_{7/2}
  +\Omega_{\gamma\delta}\Pi^{\gamma\alpha}\Pi^{\beta\delta}_{3}
  +\frac{1}{2}\Pi^\alpha_{3/2}\Pi^{\beta}_{5/2}
  +\frac{1}{2}\Omega_{\gamma\delta}\Pi^{\gamma\alpha}_{2}\Pi^{\beta\delta}_{2}.
\end{eqnarray}
The extended superalgebra ${\cal G}(4)$ is an extension of the reduced M-algebra.
\begin{eqnarray}
{\cal G}(7/2)\to {\cal G}(4)~~.
\end{eqnarray}
If it is possible to extend to a superalgebra ${\cal G}(7/2)\cdots{\cal G}(\infty)$
with suitable reducibility of generators analogous to \bref{Pie},
the asymptotic algebra is osp(1$\mid$32).  
\par

\medskip

\section{ Conclusions and discussions}\par
\indent

We proposed a generalized IW contraction.
While the original IW contraction restricts parameters contracted to 
 infinitesimally small values 
and keeps leading terms in LI one-forms,
this generalization relaxes this restriction in such a way that
not only leading terms but also next to leading terms are kept:
$L_2^z$ for the SO(3) case 
or $
L_{3/2}^\alpha
$ for the super-AdS case.
A further extension was also considered
where the guiding principle of extension is
 the MC equations expanded in parameter,
\bref{obo} for the SO(3) case 
or
\bref{g(N,M,P+1/2)}
for the super-AdS case.
A suitable  combination of generators should be considered for each contraction
in order to satisfy the cyclic identity for
 supersymmetric theories,
as well as the similarity transformation required
 even in the original IW contraction.
This contraction keeps a part of the scale parameter information 
as expected.
This is a kind of a central extension after performing the IW contraction.
In this procedure the number of independent generators of ${\cal G}(N)$ becomes infinity
at $N\to \infty$ which looks different from the number of the original algebra.
However
this is reasonable in the sense that an
infinite number of nilpotent generators is
 required to represent the
 homogeneous algebra $\ni {\bf M}$
\begin{eqnarray}
&&~~~\overbrace{\hspace{31mm}}^{N}\nn\\
&&\left(
  \begin{array}{ccccc}
      0 & 1   &  0  & \cdots  & 0   \\
      0 & 0   &  1  &\ddots   &  \vdots  \\
      0  & 0    &  0  &\ddots   &  0  \\
      \vdots  & \ddots    &  \ddots   & \ddots   & 1   \\
      0 &  \cdots   &  \cdots   & 0 & 0   
  \end{array}
\right)^{N-1}\neq 0
~\stackrel{N\to \infty}{\longrightarrow}~
{\bf M}^N={\bf M}~~.
\end{eqnarray}
Again a suitable finite number of combinations of the infinite number of generators
is required.
It is interesting that the Wess-Zumino term of the type \bref{WZ11} for a flat superstring 
is derived from homogeneous supergroup OSp(1$\mid$32)
by the usual IW contraction shown in the Chern-Simons supergravity context \cite{NMora}.

\vskip 6mm
{\bf Acknowledgments}\par
We would like to thank Michael B. Green for useful comments on the manuscript and
 Satoshi Iso for fruitful discussions.
M.S. gratefully acknowledges support by Nishina Memorial Foundation.

\end{document}